\documentclass[aps,twocolumn,showpacs,floatfix]{revtex4}

\pdfoutput=1

\usepackage{amsmath,amssymb,bm}
\usepackage{graphicx}
\usepackage{latexsym}

\def\beg{\begin{equation}}
\def\en{\end{equation}}
\def\bea{\begin{eqnarray}}
\def\eea{\end{eqnarray}}
\def\ba{\begin{array}}
\def\ea{\end{array}}

\newcommand{\bk}{{\bf k}}
\newcommand{\bx}{{\bf x}}

\newcommand{\by}{{\bf{y}}}
\newcommand{\bz}{{\bf{z}}}
\newcommand{\hx}{\hat {\bf{x}}}
\newcommand{\hy}{\hat {\bf{y}}}
\newcommand{\hz}{\hat {\bf{z}}}
\newcommand{\bR}{{\bf{R}}}
\newcommand{\pmat}[1]{\begin{pmatrix}#1\end{pmatrix}}

\newlength{\upit}\upit=0.1truein

\newcommand{\ltappr}{{{\lower4pt\hbox{$<$} } \atop \widetilde{ \ \ \ }}}
\newlength{\bxwidth}\bxwidth=1.5 truein

\newcommand{\dg}{^{\dagger }}

\begin{document}

\title{A new exotic state in an old material: A tale of SmB$_6$}
\author{Maxim Dzero}
\affiliation{Department of Physics, Kent State University, Kent, OH 44242}
\author{Victor Galitski}
\affiliation{Condensed Matter Theory Center and Department of Physics, University of
Maryland, College Park, MD 20742}
\date{\today}

\begin{abstract}
We review current theoretical and experimental efforts to identify a novel class of intermetallic $4f$ and $5f$ orbital materials in which strong interactions between itinerant and predominately localized degrees of freedom gives rise to a
bulk insulating state at low temperatures, while the surface remains metallic. This effect arises due to inversion of even 
parity conduction bands and odd parity very narrow $f$-electron bands. The number of band inversions is mainly determined by the crystal symmetry of a material and the corresponding degeneracy of the hybridized $f$-electron bands. For odd number of band inversions the metallic surface states are chiral and therefore remain robust against disorder and time-reversal invariant perturbations. We discuss a number of unresolved theoretical issues specific to topological Kondo insulators and outline experimental challenges in probing the chiral surface states in these materials.  
\end{abstract}

\maketitle

% Plan of the paper:

% What needs to be discussed:
% 1. cubic & tetragonal
% 2. holes vs. electrons
% 3. SmB6 & Ce3Bi4Pt3
% 4. Theory: bands and topologically active points in BZ
% 5. ARPES can identify the hybridization points - in SmB6 these are X so it must be STI
% 6. Speculate on other experimental consequences 

\section{Introduction}
In the past ten years researchers have been fascinated with a peculiar kind of materials: topological insulators \cite{Bernevig,Hasan2010,Qi2010,Fu2007,Moore2007,FuKane}. These materials host spin-momentum locked (i.e. chiral) metallic surface states, which allow them to remain robust to time-reversal invariant perturbations \cite{Roy2009,Hsieh2008,Xia2009,exp1,exp2,exp3,exp4}. In addition, these materials, when brought into contact with s-wave superconductors, will support Majorana  fermions that are their own antiparticles \cite{Majorana1,Majorana2}. The combination of these properties makes topological insulators promising platforms for spintronics and quantum computing applications. At the same time materials, which have been proven to possess topologically protected metallic surface states, have significant bulk conductivity \cite{Bulk1,Bulk2,Bulk3,Bulk4}. 
Therefore, in this sense they are not ideal topological insulators. 

One promising route for discovery of ideal topological insulators is to examine materials with strong electron-electron interactions. First, the electron correlations may fully suppress the bulk conductivity. Secondly, electronic interactions may significantly enhance the spin-orbit coupling, which is responsible for the inversion of the bands with opposite parity. In weakly correlated Bi-based topological insulators spin-orbit coupling inverts the $s$- and $p$-bands. In correlated topological insulators we expect bands with higher orbital number, either $p$- and $d$-orbitals or $s,d$- and $f$-orbitals to invert. For example, topological Mott insulating state has been theoretically predicted for $d$-orbital pyrochlore irridates within the extended Hubbard model on a honeycomb lattice \cite{Raghu2008,pyro1,pyro2,pyro3,pyro4}, while inversion between Os $d$-bands and Ce $f$-bands leads to topological insulator in filled skutterudites \cite{Skutter} and
general two-dimensional Kondo system where topological insulating state is hidden inside the ferromagnetic metallic state \cite{2DKondo}. It is also worth mentioning the theoretical realization of various interaction driven topological phases in ultracold atom systems and graphene \cite{Sun2009,Nandkishore2010,Sun2010}.

%Advantages of topological Kondo insulators. 
In this article, we will focus on recent theoretical and experimental breakthroughs in the search for the ideal topological insulator in higher orbital systems. The special attention has been given to already existing $f$-orbital materials \cite{KIReviews1}, such as CeNiSn, Ce$_3$Bi$_4$Pt$_3$, YbB$_{12}$ and SmB$_6$. These materials, which are called Kondo insulators, have all the necessary features needed for realizing topological behavior: strong spin-orbit coupling, strong electron-electron interactions and orbitals with opposite parity, Table I. Strong spin-orbit coupling is inherent in $f$-electron systems and it guarantees the inversion of the bands 
at the high symmetry points in the Brillouin zone (BZ). Predominantly localized character of the $f$-electrons furnishes strong Coulomb repulsion between them, while hybridization between the even parity conduction electrons and $f$-electrons leads to the emergence of the hybridization gap. Interestingly enough, in some Kondo insulators onset of the hybridization gap, observed by Raman spectroscopy, has a clear features of the second order phase transition. In any case, the opening of the hybridization gap does not guarantee an insulating gap, of course. However, in Kondo insulators total number of conduction and $f$-electrons per unit cell is even and, consequently, it immediately follows that Kondo insulators are strongly correlated analogues of band insulators. 

\begin{table}[ht] \caption{Strength of Hubbard interaction $U$ and spin-orbit coupling $\lambda_{SO}$ 
depending on a type of orbital state} 
\centering \begin{tabular}{c c c c c} 
\hline\hline  & 4$d$ & 5$d$ & 4$f$ & 5$f$ \\ [0.5ex]	% inserts table %heading 
\hline U (eV) & 1.5 & 1 & 1.7 & 2.1 \\ $\lambda_{SO}$ (eV) & 0.1 $\div$ 0.2 & 0.4 $\div$ 0.6 & 0.7$~\div$ 1 & 1$~\div$ 2\\ [1ex] 
\hline 
\end{tabular} \label{table:nonlin} 
\end{table}

This article is organized as follows. In the next Section we will review the theoretical models, which lead to the original prediction \cite{Dzero2010} that Kondo insulators with tetragonal or orthorhombic crystalline symmetries can naturally become a host to topologically protected metallic surface states. Section III will be devoted to the review of recent experimental and theoretical efforts towards the understanding the physics of cubic topological Kondo insulators and, specifically, SmB$_6$. In Section IV we discuss the open questions answer to which will deepen our understanding of topological Kondo insulators. We summarize the current status of the field and present our conclusions in Section V.
\section{theories of topological Kondo insulators}
We Section we review the recent theories of topological Kondo insulators. We consider the case when the
$f$-ion is in tetragonal crystalline field environment and review the theoretical results obtained for this case first. We then proceed
with the discussion of the theories for cubic topological Kondo insulators, which are relevant for SmB$_6$, YbB$_{12}$
and Ce$_3$Bi$_{4}$Pt$_3$ materials.  
\begin{figure}[h]
\includegraphics[scale=0.08,angle=0]{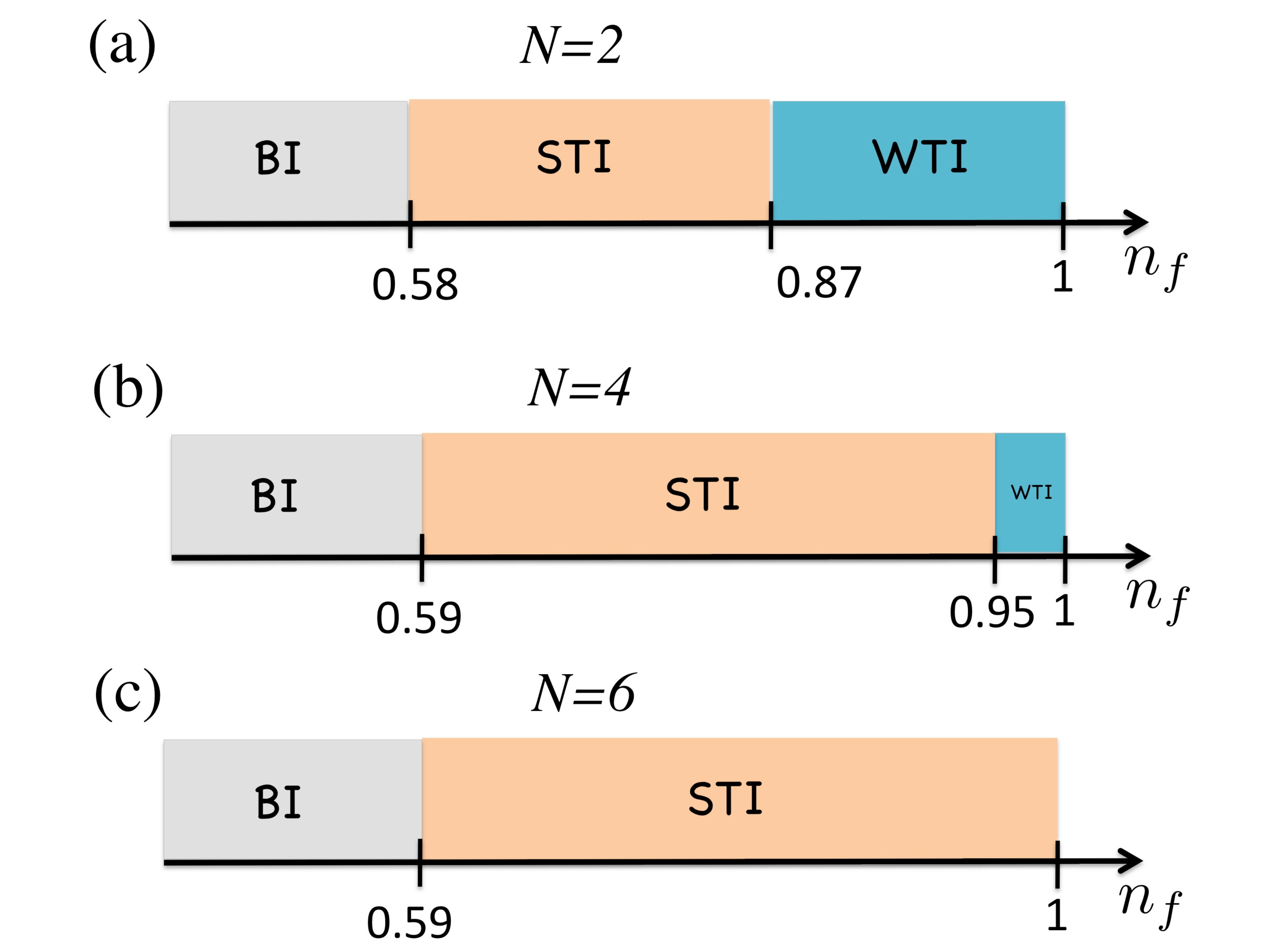}
\caption{Large-$N$ mean-field theory results for the topological Kondo insulator \cite{DzeroLargeN}. Panel (a): Phase diagram calculated for the case of doubly degenerate $f$-ion multiplet, $N=2$, as a function of the average electronic occupation of the $f$-level, $0\leq n_f\leq 1$. In the local moment regime, $n_f\approx1$, the weak topological insulator (WTI) with the topological invariant $\nu=(0;111)$ is realized. In the mixed-valence regime, $n_f\sim 1$, Kondo insulator is a strong topological insulator (STI) with the topological invariant given by $\nu=(1;111)$. When the hybridization between the conduction and $f$-electrons becomes even stronger, then material is a trivial or band insulator (BI), $\nu=(0;000)$. Panels (b) and (c): phase diagram for $N=4$ and $N=6$ correspondingly. Note that WTI state disappears as the the degeneracy of the $f$-multiplet grows.}
\label{Fig1}
\end{figure}
%%%%%% Done with Fig.1 %%%%%%%%%%%%%%%%%%%%%%%
\subsection{tetragonal topological Kondo insulators}
% Review theory: 1) PRL + PRB
% Review theory: 2) Large N approaches - Kim + Dzero
% Review theory: 3) DMFT - interaction effects
%%%%%% This is Fig. 1 -> TITLE PLOT %%%%%%
The minimal model of Kondo insulators must involve conduction and strongly correlated $f$-electrons as well as
hybridization between them. Before we write down the corresponding periodic Anderson model, we need to specify the
$f$-electron states first. Since most of the tetragonal systems which are insulating or semi-metallic contain Ce, we consider the model for the Ce ion in the total angular momentum $J=5/2$ state. The six-fold degenerate multiplet is then split into the three Kramers doublets
with the eigenvectors written conveniently in terms of the eigenvectors of the angular momentum projection operator 
$\hat{J}_z$ as follows \cite{Miyazawa2003}: 
\beg\label{doublets}
\begin{split}
|\mu=\pm1\rangle&=|\pm1/2\rangle, \\
|\mu=\pm2\rangle&=\cos\alpha|\pm5/2\rangle-\sin\alpha|\mp3/2\rangle, \\
|\mu=\pm3\rangle&=\sin\alpha|\pm5/2\rangle+\cos\alpha|\mp3/2\rangle.
\end{split}
\en
where angle $\alpha$ determines the degree of admixture between the corresponding orbitals. These eigenvectors can be 
conveniently expressed in terms of the spin part of the electron wave function $\chi_\sigma$ as follows:
\beg
|\mu\rangle=\sum\limits_{M=-5/2}^{+5/2}B_{\mu M}\sqrt{4\pi}\sum\limits_{m=-3}^3
A_{lm\sigma}^M|m,l\rangle\chi_\sigma
\en
where $A_{lm\sigma}^M$ are the Clebch-Gordon coefficients, $B_{\mu M}$ are some known constants determined from the crystalline electric field (CEF) potential.

The minimal model Hamiltonian for the tetragonal Kondo insulator then takes the following form \cite{Dzero2010,Miyazawa2003,KIReviews2}:
\beg\label{Hamiltonian}
\begin{split}
\hat{H}=&\sum\limits_{\bk\sigma}\xi_{\bk}\hat{c}_{\bk\sigma}^\dagger\hat{c}_{\bk\sigma}\\
+&\sum\limits_{\bk\mu}\varepsilon_{\bk}^{(f)}\hat{f}_{\bk\mu}^\dagger\hat{f}_{\bk\mu}+\sum\limits_{\bk\mu\sigma}
\left(V_{\bk\sigma\mu}^*\hat{f}_{\bk\mu}^\dagger\hat{c}_{\bk\sigma}+\textrm{h.c.}\right)\\
+&\frac{1}{2}U_{ff}\sum\limits_{i;\mu\not=\mu'}\hat{f}_{i\mu}^\dagger\hat{f}_{i\mu}\hat{f}_{i\mu'}^\dagger\hat{f}_{i\mu'},
\end{split}
\en
where $\hat{c}_{\bk\sigma}^\dagger$ creates an electron in the conduction band in a plane-wave state with momentum $\bk$, spin $\sigma=\uparrow,\downarrow$ and energy $\xi_{\bk}$ (relative to the chemical potential of the conduction band), 
while $\hat{f}_{\bk\mu}^\dagger$ creates an
$f$-electron in a state with momentum $\bk$ and the multiplet component $\mu$ (\ref{doublets}) and energy 
$\varepsilon_{\bk}^{(f)}$. We note that the bandwidth of the $f$-electrons is much smaller than the one for the conduction electrons. The third term (\ref{Hamiltonian}) describes the momentum dependent hybridization between the conduction and $f$-electrons, while the last terms accounts for strong local correlations between the $f$-electrons on site $i$. This last terms is important as it leads to the local moment formation. 
%%%% Fig 2 %%%%
\begin{figure}[h]
\includegraphics[scale=0.25,angle=0]{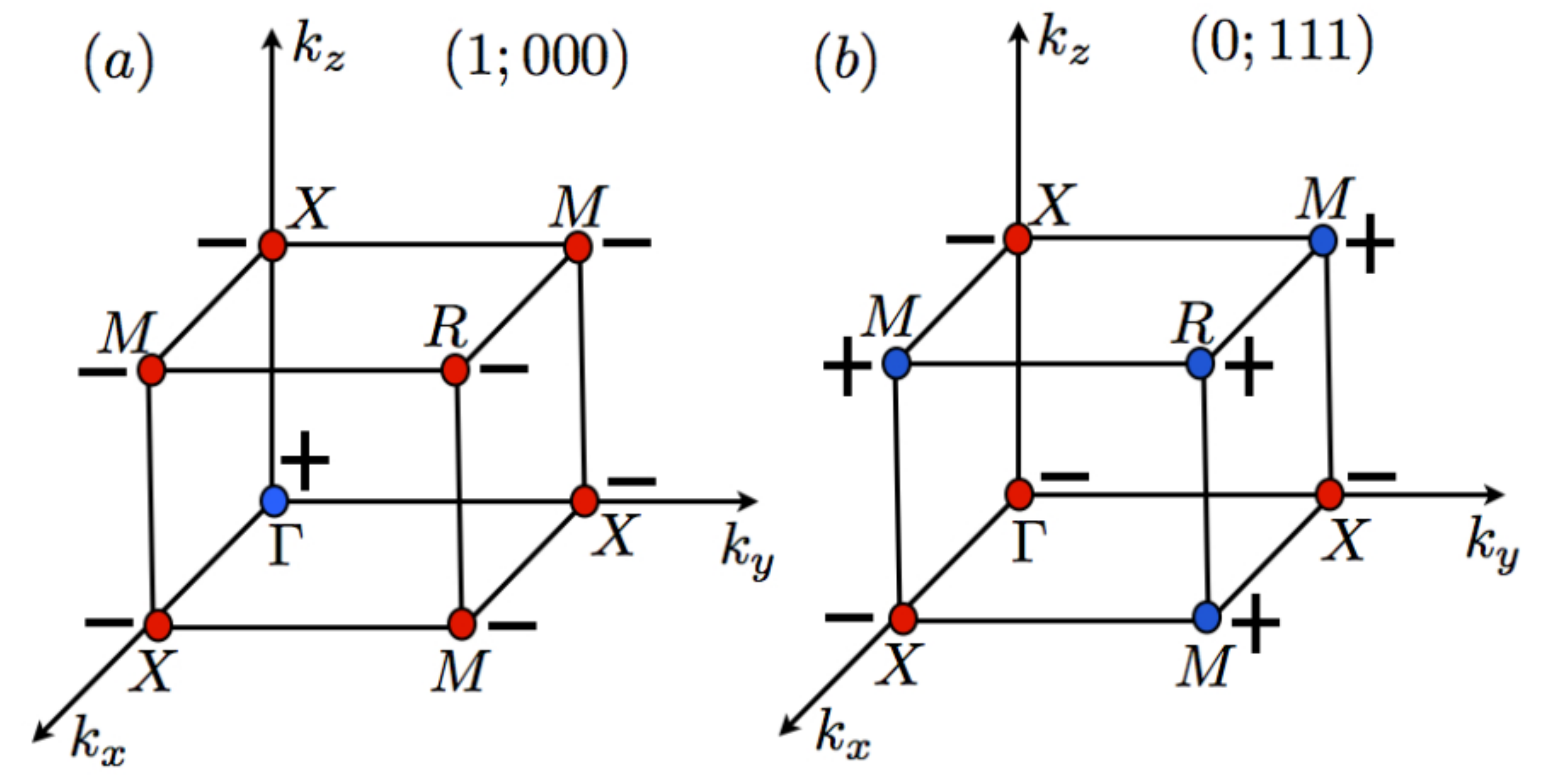}
\caption{Two topologically distinct states can be realized in our model of tetragonal topological Kondo insulators. The first one has a topological invariant $\nu=(1;000)$ and it corresponds to a strong topological insulator. The second one with topological invariant $\nu=(0;111)$.}
\label{Fig2}
\end{figure}
%%%%%% Done with Fig.2 %%%%%%%%%%%%%%%%%%%%%%%

It is very intuitive that the momentum dependence of the hybridization amplitude determines the anisotropy of the hybridization gap, which becomes an insulating gap if total number of electrons per unit cell is even. 
Formally, the momentum dependence of $V_{\bk\mu\sigma}$ can be written in terms of the spherical harmonic functions
\cite{Miyazawa2003,Dzero2012}:
\beg\label{Vk}
\begin{split}
V_{\bk\mu\sigma}=&\sum\limits_{M=-5/2}^{+5/2}B_{\mu M}\sqrt{4\pi}V_{kl}\sum\limits_{m=-3}^3
A_{lm\sigma}^M\tilde{Y}_{l}^m(\bk),
\end{split}
\en
where $V_{kl}$ are the matrix elements, which can be expressed in terms of the corresponding Slater-Koster matrix elements \cite{SlaterKoster4fs}.  Note that the values of the momentum in (\ref{Vk})
are defined everywhere in the Brillouin zone and 
\beg
\tilde{Y}_{l}^m (\bk )= \frac{1}{Z}\sum_{\bf R \neq 0} Y_{l}^m (\hat {\bf R}) e^{i \bk\cdot {\bf R}}
\en
is a tight-binding generalization of the spherical Harmonics that
preserves the translational symmetry of the hybridization,
$V_{\bk}= V_{\bk +{\bf G}}$,  where ${\bf G}$ is 
reciprocal lattice vector \cite{Dzero2010,Dzero2012}.  Here, 
 $\bf R$ are the positions of the Z nearest neighbor sites
around the magnetic ion. 

The low energy properties of the model (\ref{Hamiltonian}) can be analyzed by the employing the following conjecture: the effect of the local correlations between the $f$-electrons leads to the renormalization of the hybridization amplitude
and the shift of the $f$-energy level:
\beg\label{lowEmodel}
\begin{split}
&\varepsilon_{\bk\mu}^{(f)}\to\tilde{\varepsilon}_{\bk\mu}^{(f)}=Z_{\bk\mu}\left(\varepsilon_{\bk\mu}^{(f)}+\Sigma_{\mu\mu}(\bk,\omega=0)\right), \\
&|V_{\bk\mu\sigma}|\to|\widetilde{V}_{\bk\mu\sigma}|=\sqrt{Z_{\bk\mu}}|V_{\bk\mu\sigma}|,
\end{split}
\en
where the renormalization factor $Z_{\bk\mu}$ is determined by the $f$-electron self-energy part 
$\Sigma_{\mu\mu}(\bk,\omega)$:
\beg\label{Zk}
Z_{\bk\mu}=\left[1-\frac{\partial \Sigma_{\mu\mu}(\bk,\omega)}{\partial \omega}\right]_{\omega=0}^{-1}.
\en
Then the low-energy model can be diagonalized yielding a band structure which consists of two doubly degenerate bands separated by the momentum dependent energy gap $\Delta_\bk$ given by 
\beg\label{Delta}
\Delta_{\bk}^2=\frac{1}{2}\textrm{Tr}[V_{\bk}^\dagger V_{\bk}],
\en
where we suppressed the spin and orbital indices for brevity.
Since the total number of electrons per unit cell is even, we are guaranteed to have the lowest two bands to be fully occupied and we have a band insulator provided that $\Delta_{\bk}$ does not have nodes anywhere in the BZ, apart from the high-symmetry points, where it vanishes since $V_{-\bk;\mu\sigma}=-V_{\bk,\mu,\sigma}$. Analysis of the momentum dependence of the function $\Delta_{\bk}$ shows that for $\mu=\pm2,\pm3$, function $\Delta_{\bk}$ vanishes at $k_z=\pm k$, i.e. gap has two point nodes \cite{Miyazawa2003}. Therefore, hybridization and insulating gap always open for $\mu=\pm1/2$, corresponding to a ground state multiplet with $J_z=\pm1/2$. In passing, we note that for systems with lower symmetry (i.e. orthorhombic), the ground state Kramers doublet will be given by linear combination of states (\ref{doublets}). In this case, despite the fact that $J_z=\pm1/2$ is nodeless, one can show that hybridization gap acquires nodes at some finite values of momentum $\bk$ (see discussion in Ref. \cite{Miyazawa2003} for details). 
% Still need to discuss:
% 1. Topological invariant

The parity at a high symmetry point can be determined by $\delta_m=\textrm{sgn}(\xi_{\bk^*_m}-\tilde\varepsilon_{\bk^*_m}^{(f)})$ \cite{Bernevig,Dzero2010,Dzero2012,FuKane}.
Four independent $Z_2$ topological indices $(\nu_0;\nu_1\nu_2\nu_3)$ ~\cite{Kitaev}, one strong ($a=0$) and three weak indices ($a={1,2,3}$) can be constructed from $\delta_m$: (i)~The strong topological index is the product of all eight $\delta_m$'s: $I_{\rm STI} = (-1)^{\nu_0}=\prod\limits_{m=1}^{8} \delta_m = \pm 1$; 
(ii)~by setting $k_j=0$ (where $j= x,y, \mbox{and } z$),  
three high-symmetry planes, $P_j = \left\{ {\bf k}: k_j=0\right\}$, are formed that contain four high-symmetry points each. The product of the parities at these four points defines the
corresponding weak-topological index, $I_{\rm WTI}^a =(-1)^{\nu_a}= \prod\limits_{{\bf k}_m \in P_j} \delta_m = \pm 1$, $a=1,2,3$ with integers corresponding to the axes $x,y$ and $z$. The existence of the three weak topological indices in 3D is related to a $Z_2$ topological index for 2D systems (a weak 3D TI is similar to a stack of 2D $Z_2$ topological insulators). Because there are three independent ways to stack 2D layers to form a 3D system,
the number of independent weak topological indices is also three.
A conventional band insulator has all of the four indices $I_{\rm STI}
= I_{\rm WTI}^x=I_{\rm WTI}^y=I_{\rm WTI}^z = +1$ or  equivalently (0;000). An index
$I=(-1)$ ($\nu_a=1$) indicates a $Z_2$ topological state with the odd number of
surface Dirac modes. In tetragonal KI  inversion index $\delta_{m}$ of a particular
high symmetry point $m$ is negative if the conduction band is below the $f$-band: 
$\xi_{\bk^*_{m}}<\tilde\varepsilon_{\bk_{m}^*}^{(f)}$. 
Thus if $\xi_{{\bk_m^*}=0}<\tilde\varepsilon_{{\bk_m^*}=0}^{(f)}$  at the $\Gamma$ point, while the remaining high symmetry points remain inert, $\xi_{\bf{{\bk^*_m\ne 0}}}>\tilde\varepsilon_{\bk_m^*}^{(f)}$, then $I_{\rm STI}= -1$, and hence the tetragonal Kondo insulator is a strong-topological insulator, robust against disorder. Fig. \ref{Fig2}(a). 
Weak-topological insulators and topologically trivial
insulators can in principle be found for different band structures and
different values of $\tilde\varepsilon_{\bk_m^*=0}^{(f)}$, Fig. \ref{Fig2}(b). 

% 2. Large N
One can go beyond the phenomenological description described above and resort to a more microscopic approach. 
Specifically, we can consider the limit of $U_{ff}\to\infty$ and work in the restricted phase space by projecting out the doubly occupied $f$-states. Formally, this is accomplished by introducing the slave boson operators \cite{Slave1,Slave2,Slave3,Slave4,Slave5}. Then, the mean-field analysis of the resulting model can be made by replacing the slave boson operators by a number, which must be determined self-consistently. The mean-field theory is controlled by the degeneracy of the $f$-orbital multiplet. The large-$N$ mean-field analysis for the case of tetragonal Kondo insulator have been done by Tran, Takimoto and Kim \cite{Tran2012} and also by Dzero \cite{DzeroLargeN}. The results of the slave-boson mean-field theory generally agree with the phenomenological approach for $N=2$, Fig. \ref{Fig1}(a). In Ref. \cite{DzeroLargeN} the mean-field diagram has also been obtained for $N>2$, Fig. \ref{Fig1}(b,c). Interestingly, it was found that the WTI state is suppressed when the degeneracy of the $f$-multiplet grows, so for $N=6$ only STI state is present. This result indicates that STI state should be preferred for higher symmetry, i.e. cubic, systems. We will return to this observation when we discuss the cubic topological Kondo insulators below. 
%%%%%% This is Fig. DMFT -> TITLE PLOT %%%%%%
\begin{figure}[h]
\includegraphics[scale=0.08,angle=0]{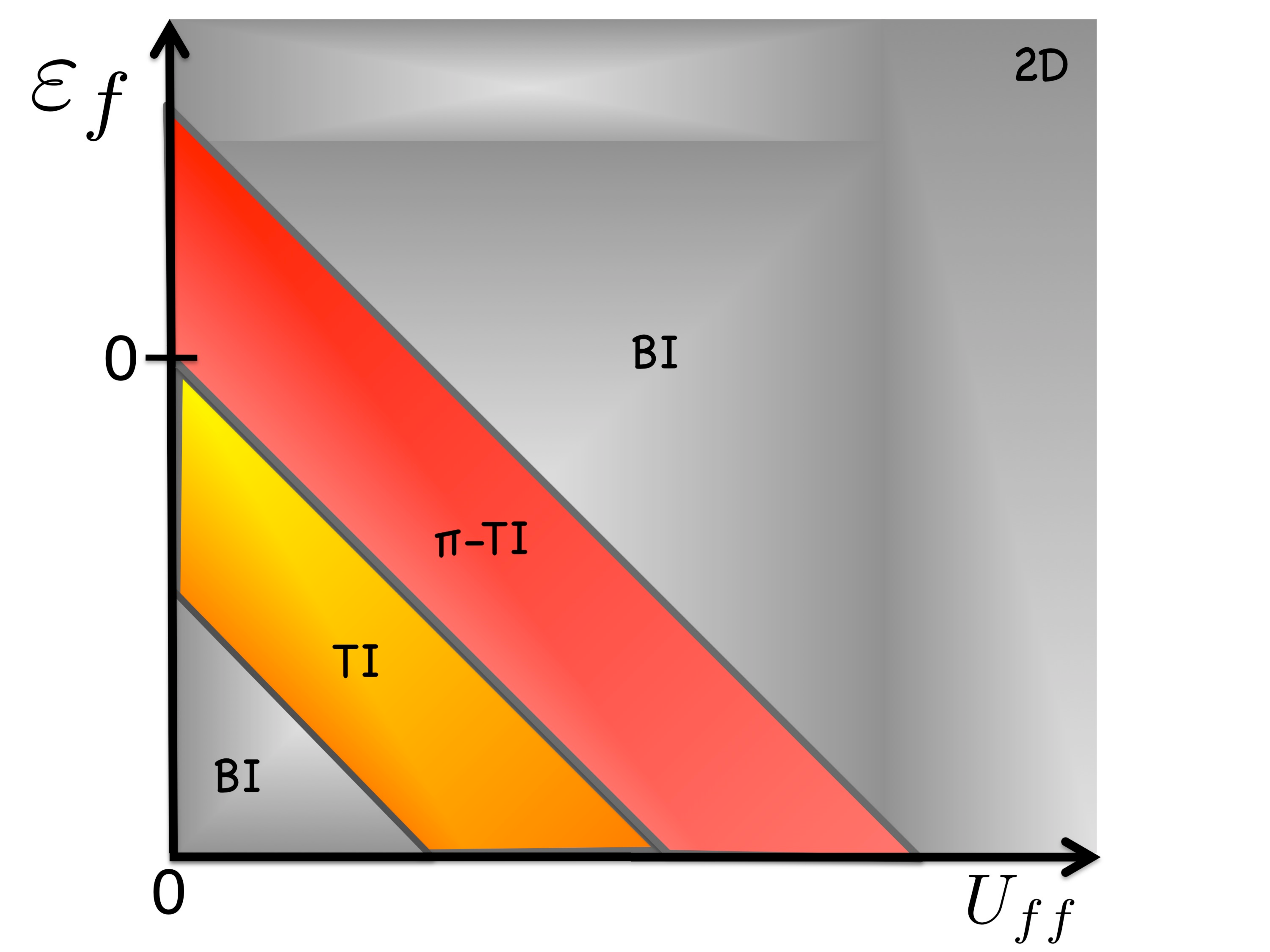}
\caption{Dynamical mean field theory phase diagram for the two-dimensional (2D) Kondo insulator (adopted from Ref. \cite{DMFT}).
As the strength of the Hubbard interaction $U_{ff}$ between the $f$-electrons increases, there are three topologically distinct insulating states. Band insulator (BI) is realized for small (compared to the conduction electron bandwidth) values of $U_{ff}$. Topological
insulator (TI) with band inversion at the $\Gamma$ point sets in for intermediate values of $U_{ff}$. When the interaction strength $U_{ff}$ increases further, this state is succeeded by a topologically non-trivial $\pi$-TI state: in this state bands of the opposite parity invert at the X points of the 2D Brillouin zone. At the boundaries between topologically distinct states, the system is semi-metallic.}
\label{FigDMFT}
\end{figure}
%%%%%% Done with Fig.DMFT %%%%%%%%%%%%%%%%%%%%%%%

% 3. DMFT: we need a schematic figure
The phenomenological and mean-field models discussed so far assume that the interaction between $f$-electrons
is infinitely large. How will the result change if we consider the finite values of $U_{ff}$? Can one study the evolution
of the topological state as a function of $U_{ff}$? In other words, can Kondo insulator be adiabatically connected to 
a non-interaction band insulator? This and related questions has been recently addressed by Werner and Assaad \cite{DMFT} by using the dynamical mean-field theory (DMFT) to analyze the two-dimensional Kondo insulators.

The physical properties of the model described by the Anderson lattice Hamiltonian (\ref{Hamiltonian}) can be captured, in particular, by the single particle propagators. The self-energy corrections in the single particle correlation functions are governed
by the Hubbard repulsion between the $f$-electrons. Generally, the $f$-electron self-energy part depends both on momentum and frequency. Within the DMFT, however, one considers a single $f$-electron site version of the lattice model (\ref{Hamiltonian}) lattice, which allows one to compute the single particle and higher order correlation functions exactly \cite{Kristian}. Consequently, the self energy parts, which encode the correlation between the $f$-electrons, are all momentum independent. Then, the self energy parts for the conduction and $f$-electrons corresponding to the lattice problem can be determined self-consistently from the solution of the impurity problem. 

Interestingly Werner and Assaad have found \cite{DMFT} that increasing the strength of the Hubbard interaction leads to the series of transitions between normal insulator (small $U_{ff}$) into a strong topological insulator with the in gap state crossing the Fermi level at the $\Gamma$ point and then into another strong topological insulator, in which the in-gap states cross the Fermi level at the X point, Fig. \ref{FigDMFT}. These results clearly indicate that previously developed concept of adiabatic connection between uncorrelated band insulators and strongly correlated Kondo insulators \cite{Varma1977} does not hold at least for the two-dimensional Kondo lattice. It remains to be verified, of course, whether this conclusion will holds for 3D Kondo insulator. In conclusion, we note that true tetragonal Kondo insulators still remain to be experimentally discovered, since until now Ce-based tetragonal "Kondo insulators", such as CeNiSn \cite{cenisn} and
CeRuSn$_6$ \cite{cerusn6}, upon improving sample's quality become semi-metals. 

\subsection{cubic topological Kondo insulators}
% All currently known Kondo insulators are cubic. 
% Review of band structure calculations: LDA & LDA + Gutzwiller: main conclusion - X points
% Mean-field theories: bands invert @ X points 
% SmB6, YbB12 - hole materials; Ce3Bi4Pt3 - cubic
Kondo insulators which consistently show insulating behavior in transport measurements - most notably SmB$_6$, YbB$_{12}$ and Ce$_3$Bi$_4$Pt$_3$ - are all cubic. In this section we will review the recent theories of cubic topological Kondo insulators, using samarium hexaboride as a specific example \cite{Takimoto2011,Dai2012,Alexandrov2013}. We note, however, that the model we will be discussing below should also hold for Ce$_3$Bi$_4$Pt$_3$ although the some parameters can be different. 

The magnetic valence configuration of the Sm ion corresponds to the state with the total angular momentum $J=5/2$. The six-fold
degenerate Sm multiplet is split by cubic crystal fields into the $\Gamma_7$ Kramers doublet and $\Gamma_8$ quartet. Consequently 
the eigenstates of the cubic crystalline field Hamiltonian are given by:
\beg
\begin{split}
|\Gamma_7,\pm\rangle&=\sqrt{\frac{1}{6}}|\pm5/2\rangle-\sqrt{\frac{5}{6}}|\mp 3/2\rangle, \\
|\Gamma_8^{(1)},\pm\rangle&=\sqrt{\frac{5}{6}}|\pm5/2\rangle+\sqrt{\frac{1}{6}}|\mp 3/2\rangle, \\
|\Gamma_8^{(2)},\pm\rangle&=|\pm1/2\rangle.
\end{split}
\en
If $\Gamma_7$ is the ground state multiplet the system is a semimetal, since the hybridization gap will have nodes in the BZ \cite{Miyazawa2003}. Thus, for the insulating state one necessarily needs $\Gamma_8$ to be the ground state multiplet. This appears to be indeed the case for SmB$_6$ (as well as for Ce$_3$Bi$_4$Pt$_3$) as evidenced by inelastic neutron scattering experiments and Raman spectroscopy \cite{neutronsSmB61,neutronsSmB62,Raman2,Raman1}.

\begin{figure}[h]
\includegraphics[scale=0.07,angle=0]{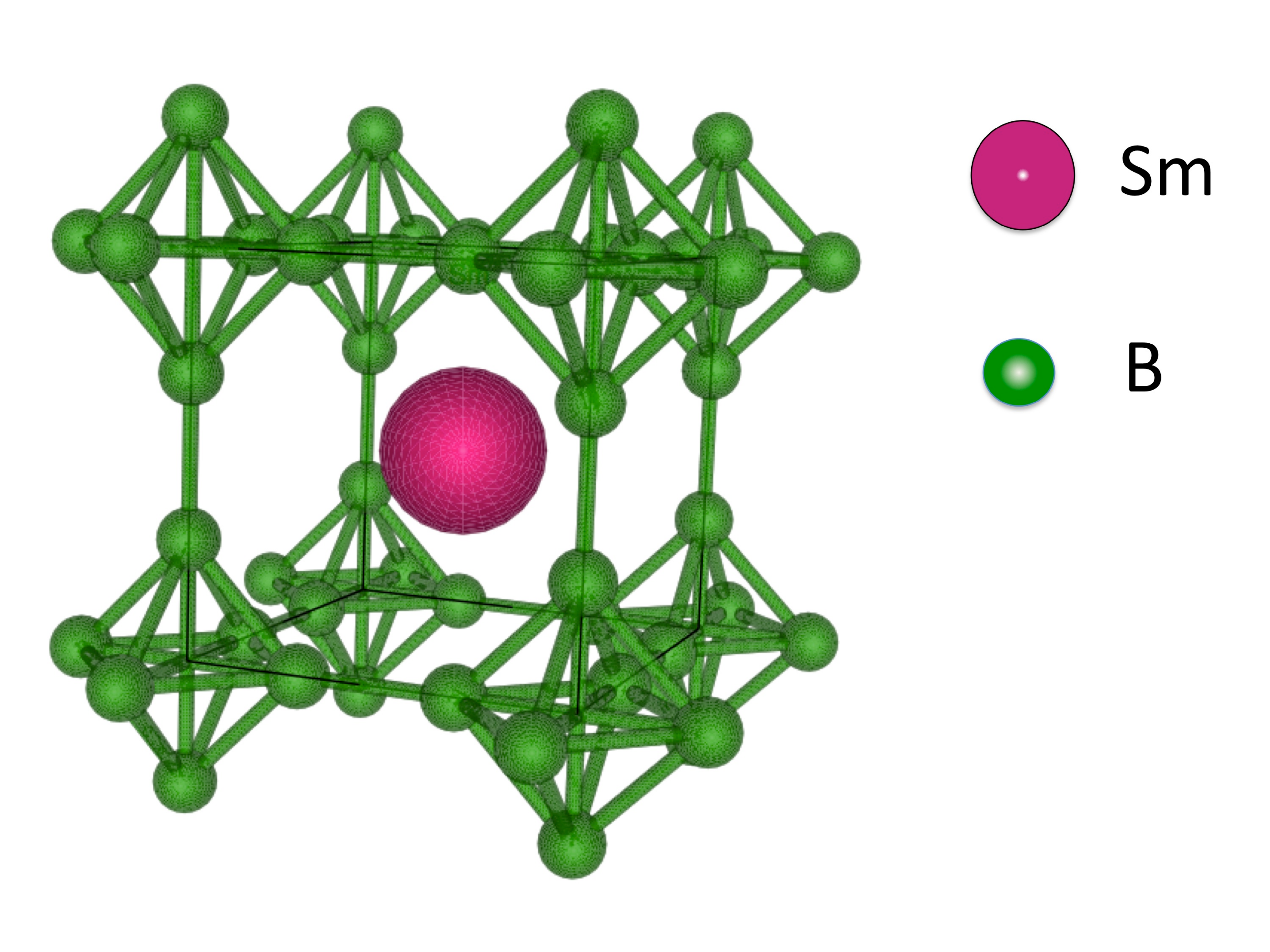}
\caption{Crystal structure of SmB$_6$. Sm ions are at the center of the unit cell and are surrounded by octahedrons of boron ions located at the corners of the unit cell.}
\label{FigSmB6}
\end{figure}

SmB$_6$ has a cubic CsCl-like crystal structure, Fig. \ref{FigSmB6}, with the B$_{6}$ clusters
located at the center of the unit cell. From band-theory calculations \cite{Dai2012,Antonov}, the B$_{6}$ clusters act as spacers which mediate  electron hopping between Sm sites, but are otherwise inert. In addition to band structure results, x-ray photoemission spectroscopy (XPS) of SmB$_6$ \cite{Raman2} indicate that the conduction bands that hybridize with the localized 4$f$-orbitals
are 5$d$-states which form electron pockets around the X points. In particular, the physics of the $4f$ orbitals is governed by valence fluctuations involving electrons of the $\Gamma_{8}$ quartet and the conduction $e_g$-hole states, 
$4f^{5}\rightleftharpoons4f^{6}+h$. The conduction states must be $d_{x^2-y^2}$ and $d_{3z^2-r^2}$ orbitals of $e_g$ symmetry. Since the lowest lying state must be a quartet, one immediately concludes that the band inversion happens odd number of times and cubic Kondo insulator must be strong topological insulator for a moderate value of the hybridization \cite{Alexandrov2013}. Indeed, at the $\Gamma$ point both $d$ and $f$-bands remain four-fold degenerate, so that $\Gamma$ point remains topologically inert. 

Recently, Takimoto \cite{Takimoto2011} has performed the low-energy analysis of the Anderson Kondo lattice model (\ref{Hamiltonian}) 
properly generalized to take into account the realistic band structure of SmB$_6$. Specifically, the conduction and $f$-electron spectrum have been derived using the tight-binding approximation to the next-nearest-neighbors approximation. It is important to emphasize that next-nearest-neighbors hopping is needed in order to obtain the minimum of the conduction bands at the X points of the BZ. The hybridization part of the Hamiltonian has also been derived using the tight-binding approximation restricted to the nearest-neighbors only. One can justify 
this approximation by noting that the orbital momentum $\Delta l=\pm1$ is transferred from the conduction $d$-states to the $f$-states in the process of hybridization. The phenomenological analysis similar to the one of Ref. \cite{Dzero2010} of the resulting model 
shows that SmB$_6$ is a strong topological insulator. An important conclusion one draws from the results of Ref. \cite{Takimoto2011} is that as a consequence of the band inversion at the X points, there will be three Dirac cones on the surfaces perpendicular to the main symmetry axes. Subsequent first-principles study based on the local-density-approximation (LDA) plus Gutzwiller method \cite{Dai2012} has confirmed the results of the phenomenological theory \cite{Takimoto2011}.

Most recently, Alexandrov, Dzero and Coleman (ADC) have formulated a general model for the cubic topological Kondo insulators 
\cite{Alexandrov2013}. The ADC model for the three dimensional Kondo insulator can be derived from an effective ($U_{ff}\to\infty$) model for the one-dimensional Kondo insulator by applying a series of unitary transformations. Specifically one considers  the quartet of $f$- and $d$-holes described by and orbital and spin index, denoted by the combination 
$\lambda\equiv(a,\sigma )$ ($a=1,2$, $\sigma = \pm 1$). Consequently, the $d$- and $f$-states are then described by the eight component spinor 
\begin{equation}\label{}
\Psi_{j} = \left(\begin{matrix} d_{\lambda} (j)\cr X_{0\lambda } (j)
\end{matrix} \right)
\end{equation}
where $d_{\lambda} (j)$ destroys an $d$-hole at site $j$, while
$X_{0\lambda} (j)=\vert 4f^{6}\rangle \langle  4f^{5},\lambda\vert$ is the
the Hubbard operator that destroys an $f$-hole at site $j$.
The tight-binding Hamiltonian describing the hybridized $f$-$d$ system is then
\begin{equation}\label{ADC}
H = \sum_{i,j}\Psi^\dagger_{\lambda} (i)h_{\lambda\lambda'} ({\mathbf R}_{i}-{\mathbf R}_j) \Psi_{\lambda'} (j)
\end{equation}
in which the nearest hopping matrix has the structure
\begin{equation}\label{1d3d}
h ({\mathbf R})= \begin{pmatrix} h^{d} ({\mathbf R}) & V ({\mathbf R})\cr V^\dagger ({\mathbf R})& h^{f} ({\mathbf R})
\end{pmatrix},
\end{equation}
where the diagonal elements describe hopping within the $d$- and $f$-
quartets while the off-diagonal parts describe the hybridization
between them and ${\mathbf R} \in (\pm \hx ,\pm \hy, \pm \hz)$ is the vector
linking nearest neighbors.  The various matrix elements simplify for
hopping along the z-axis, where they become orbitally and spin
diagonal:
\begin{equation}\label{}
h^{l}(\bz) = t^{l}\pmat{1 & \cr & \eta_{l}}, \qquad
V (\bz) =i  V \pmat{0 & \cr & \sigma_{z}}.
\end{equation}
where $l=d,f$ and $\eta_{l} $ is the ratio of orbital hopping elements.
In the above, the overlap between the $\Gamma_{8}^{(1)}$ orbitals,
which extend perpendicular to the z-axis is
neglected, since the hybridization is dominated by the overlap of the
the $\Gamma_{8}^{(2)}$ orbitals, which extend out along the z-axis.
%%%%%%%%
%%%%%%%%3-Dimensional Model
%%%%%%%%
The hopping matrix elements in the  $\bx $ and $\by$
directions are then obtained by rotations in orbital/spin space.
so that
$h (\bx )= U_{y}h (\bz)U_{y}\dg $ and $h (\by)= U_{-x}h (\bz
)U\dg_{-x}$ where $U_{y}$ and $U_{-x}$ denote 90$^{\circ }$ rotations
about the y and negative x axes, respectively. By construction, the hopping terms $h^{d,f}(\bk)$ in the Hamiltonian 
(\ref{ADC}) could have been alternatively obtained from the tight-binding approximation within the nearest-neighbor
approximation. However it is important to keep in mind, that the correct band structure for SmB$_6$ is still recovered
by the proper choice of the ratio between the corresponding hopping amplitudes \cite{Alexandrov2013}.

Furthermore, the Fourier transformed hopping matrices $h (\bk )= \sum_{\bR} h (\bR)
e^{-i \bk \cdot\bR}$ can then be written in the compact form
\begin{equation}\label{Hdf3D}
h^{l} (\bk )= t^{l}\pmat{\phi_{1} (\bk )+ \eta_l\phi_{2} (\bk ) & (1-\eta_{l} )\phi_{3} (\bk )\cr
(1-\eta_{l} )\phi_{3} (\bk ) & \phi_{2} (\bk ) + \eta_{l} \phi_{1} (\bk )
}+ \epsilon^{l},
\end{equation}
where $l=d,f$. Here $\epsilon^{l}$ are the bare energies of the
isolated d and f-quartets, while
$\phi_{1} (\bk)= c_{x}+c_{y}+4 c_{z}$, $\phi_2(\bk) =
3(c_{x}+c_{y})$ and $\phi_{3} (\bk )= \sqrt{3}(c_{x}-c_{y})$ ($c_{\alpha}=\cos
k_{\alpha}, \alpha=x,y,z$). The hybridization is given by
\begin{equation}\label{Hv3D}
V (\bk )=  \frac{V}{6}
\pmat{
   3 (\bar\sigma_x  +i \bar\sigma_y  )& \sqrt3( \bar\sigma_x  -i \bar\sigma_y  ) \cr
   \sqrt3( \bar\sigma_x  -i \bar\sigma_y  ) &\bar \sigma_x + i\bar \sigma_y  + 4 \bar\sigma_z  \cr
}
\end{equation}
where we denote $\bar\sigma_\alpha = \sigma_\alpha \sin k_{\alpha}$.
Naturally the hybridization between the even parity $d$-states and odd-parity $f$-states is an odd
parity function of momentum $V (\bk )=-V (-\bk )$. 

The model (\ref{ADC}) has been analyzed using the slave-boson mean-field approximation and it was shown that
cubic Kondo insulator is a strong topological insulator (STI) in complete agreement with previous works \cite{Takimoto2011,Dai2012}. 
What is more, ADC have shown that the STI state extends well into the local 
moment regime, similarly to the results form the large-$N$ theory for tetragonal Kondo insulators, Fig. 1(c). 
Lastly, we would like to emphasize that the procedure for obtaining the effective model for a cubic Kondo insulator \cite{Alexandrov2013} can be easily generalized to various type of conduction orbitals and therefore  can be used to analyze the topological properties
of the band structure for other cubic Kondo insulators such as YbB$_{12}$ and Ce$_3$Bi$_4$Pt$_3$.

\section{experiment: samarium hexaboride}
In this Section we present a brief overview of experiments on the canonical Kondo insulator
SmB$_6$ \cite{Geballe1969}. There exists a vast amount of experimental literature on this material, which merits a separate review paper. Here we will focus on discussing the experimental properties of importance for possible realization of topologically protected chiral surface states. 
% transport
% pressure
% magnetic field response
% Raman spectroscopy & neutrons
% recent experiments
%%%%%% This is Fig. 1 -> TITLE PLOT %%%%%%
%%%%%% Done with Fig.1 %%%%%%%%%%%%%%%%%%%%%%%

SmB$_6$  is the first known and most experimentally studied heavy-fermion semiconductor \cite{Geballe1969}. At high temperatures SmB$_6$ is a metal with the magnetic susceptibility showing the Curie-Weiss like behavior, $\chi\sim 1/T$ signaling the existence of the local magnetic moments originating from Sm $f$-electrons. At low temperatures, $T\ll T_{coh}\sim 50$ K, SmB$_6$ is a narrow gap semiconductor with weakly temperature dependent magnetic susceptibility \cite{KIReviews1,KIReviews2}. What is more, the average electronic occupation of the Sm $f$ states is non-integer and is varying between the Sm$^{2+}$ ($4f^6$) and Sm$^{3+}$ ($4f^55d$) configurations. This mixed valence state appears as a result of strong hybridization between the $5d$ and $4f$ Sm electrons \cite{Antonov}, which also leads to formation of the hybridization gap $E_h\approx 10$ meV and many-body insulating gap at $E_g\simeq 5$ meV. 

Most importantly, the saturation of resistivity at temperatures below $T^*\simeq5$K was consistently observed independently by many groups \cite{Varma1977}. Resistivity saturation was initially interpreted as an extrinsic effect due to formation of the impurity band. However, almost ten years after initial discovery \cite{Geballe1969},  Allen, Batlogg and Wachter \cite{AllenWachter1979} on the basis of Hall coefficient data have argued that observed values of conductivity were too small for a metallic-like system.  Moreover, with subsequent accumulation of the experimental data it became clear that this is an intrinsic electronic effect, although the controversy regarding intrinsic or extrinsic origin of the effect remained. This controversy was mainly resolved with pressure experiments 
\cite{Pressure,PressureExp1}, which have shown that for pressure above $p^*\simeq 45$ kbar the SmB$_6$ recovers its metallic properties. Interestingly, it was observed \cite{Pressure} the states contributing to low-T conductivity dominate the transport up to high enough temperatures when the transport gap becomes fully suppressed, which means that these states must be intrinsic property of a system. Transport measurements under pressure and applied magnetic field as large as $18$ Tesla have shown the negative magnetoresistance $[\Delta\rho(H)/\rho(H=0)]_{p<p^*}\sim-H^2$ at small pressure, while for pressures above $p^*$ the magnetoresistance becomes positive and grows as  $[\Delta\rho(H)/\rho(H=0)]_{p>p^*}\sim H^{3/2}$. The activation gap, on the other hand, is very weakly dependent on the magnetic field, which indicates, albeit indirectly, that the $\Gamma_8$ quartet is the ground state multiplet for Sm$^{3+}$ ions, since it has the smallest value of the $g$-factor, $g_{\Gamma_8}=2/7$ \cite{HighFields}. 

Since the seminal paper \cite{AllenWachter1979} the puzzle of temperature independent resistivity below $T^*$ became clearly recognized and its origin remained mysterious for almost thirty years. It nevertheless was widely believed that  low-temperature conductivity in SmB$_6$ originates primarily from the bulk \cite{Yanase1992,SelfTrapping,PressureHall,PressureEffectTheory,NMR} and the surface provides significantly small, if any, contribution to conductivity. From our current perspective this is quite remarkable, since the first topological insulator could have been discovered almost 30 years ago well before the discovery of the integer quantum Hall effect! Nevertheless it took many years until very recently to ask the question whether the saturation of resistivity below 5 K in SmB$_6$ is purely surface effect. Motivated by the theoretical work \cite{Dzero2010},  several groups independently \cite{exp1smb6,exp2smb6,exp3smb6} have addressed the issue of bulk vs. surface conductivity.  The transport \cite {exp1smb6}, Hall effect \cite{exp2smb6}, tunneling \cite{exp3smb6} measurements and the angular-resolved photoemission spectroscopy \cite{ARPES} unambiguously showed that below $T^*$ only surface of SmB$_6$ is conducting. What is more, the surface metallic states remains surprisingly robust against variations in the surface quality. 

Thus, the results of the most recent experiments are manifestly in support of the initial prediction \cite{Dzero2010} 
that Kondo insulators host topologically protected surface states. Indeed, robustness of the surface states, the fact that their appearance is correlated with the emergence of the hybridization gap indicate that SmB$_6$ is a strong topological insulator \cite{Takimoto2011,Dai2012,Alexandrov2013}. However, a combination of high resolution  ARPES and tunneling data are needed to directly confirm the chirality of the metallic surface states in SmB$_6$.

\section{Open questions}
% ARPES & STM - not enough resolution.
% What are the signatures of the chiral states in SmB6: de Haas - van Alfen, ultrasound, phonons etc
% Band structure calculations are needed for both YbB12 and Ce343
The theoretical and experimental results reviewed by us so far bring up a number of issues which need to be understood. For example, the Raman spectroscopy data \cite{Raman2,Raman1} seem to indicate that opening of the hybridization gap in SmB$_6$ happens in a way very similar to the second order phase transition, i.e. hybridization gap plays a role
of the mean-field order parameter. Typically mean-field-like transitions assume well defined separation of energy scales, just like it happens in conventional superconductors, for example when the small ratio of the Debye frequency to the Fermi energy renders the mean-field BCS theory to be extremely reliable. In addition, the mean-field-like onset of the hybridization implies that the fluctuations
in Kondo insulators are much weaker compared with the metallic heavy-fermion systems, which brings up the need for better understanding of the fluctuations in heavy-fermion systems. 

Theoretically, there is still an open problem of the strong coupling description of the topological Kondo insulators, similar to the 
Nozi\`{e}res Fermi liquid picture of the Kondo ground state \cite{Nozieres}. The progress towards the solution of that problem
will significantly deepen our understanding of the microscopic structure of the chiral states on the surface of the topological Kondo
insulators. 

Finally, another set of questions questions concerns Ce-based heavy fermion semiconductors which order anti-ferromagnetically. 
This materials have tetragonal crystal structure and it remains to be seen whether antiferromagnetic interactions promote
strong topological insulating state. The same applies to the possibility for topological states in heavy-fermion superconductors.

\section{Conclusions and outlook}
As the search for an ideal topological insulator continues, we have come to realizion that
at least one ideal topological insulator - samarium hexaboride - has been discovered almost 30 years ago. Current theories of topological Kondo insulators all show that the existence of the chiral surface states in $f$-orbital semiconductors is one of its fundamental signatures. Moreover, these states exist for the broad range of system's parameters, such as position of the $f$-electron chemical potential and the strength of the Hubbard interaction between the $f$-electrons. For cubic topological Kondo insulators the only requirement is that the hybridization between the conduction and the $f$-electrons must be strongest at the X or M high-symmetry points in the Brillouin zone. This guarantees the odd number of the band inversions and hence a strong topological insulator. 

% topological Kondo insulators are very promising candidates for ideal topo insulators.
In recent years we have witnessed a resurgence of theoretical and experimental activity in the field
of heavy-fermions semiconductors and there remains little doubt that SmB$_6$ is only the first topological Kondo insulators. The new potential candidates are YbB$_{12}$ and Ce$_3$Bi$_4$Pt$_3$ - materials with physical propeties
very similar to the ones of SmB$_6$. Topological Kondo insulators 
would present an ideal platform for in-depth transport studies
of chiral surface states. In addition, an interplay between strong spin-orbit coupling and electron-electron correlations
may open a way to study the broader range of effects related to the nontrivial topological structure of the electronic
states in these materials. One possible direction is the in search for topologically non-trivial states in Kondo semimetals. 
However, the most important challenge lies in developing new ways to probe the chirality of the surface metallic states whether by spin-polarized tunneling spectroscopy, by Kerr effect measurements or by radio-frequency and microwave spectroscopy. 

\begin{acknowledgments}
We are grateful to V. Alexandrov, P. Coleman, J. P. Paglione and K. Sun for discussions and collaborations on
the problems discussed in this paper. This work was financially supported by the Ohio Board of Regents Research
Incentive Program grant OBR-RIP-220573, Kent State University and the U.S. National Science Foundation I2CAM International Materials Institute Award, Grant DMR-0844115 (M.D.)  and by DOE-BES DESC000191 (V.G.).
\end{acknowledgments}

\end{document}